\documentclass{ifacconf}
\usepackage{natbib}        
\usepackage{myStyle}

\DeclareDocumentCommand{\figRef}{m}{Fig.~\ref{#1}}
\DeclareDocumentCommand{\assRef}{m}{Assumption.~\ref{#1}}

\DeclareDocumentCommand{\secRef}{m}{Section~\ref{#1}}

\DeclareDocumentCommand{\defRef}{m}{Definition~\ref{#1}}
\DeclareDocumentCommand{\exampleRef}{m}{Example~\ref{#1}}
\usepackage{url}
\usepackage{booktabs}
\usepackage{cases}

\usepackage{eso-pic}
\AddToShipoutPictureBG*{%
	\AtPageUpperLeft{%
		\setlength\unitlength{1in}%
		\hspace*{\dimexpr0.5\paperwidth\relax} 
		\makebox(0,-1.75)[c]{
			\begin{tabular}{c c}
					Max van Haren,
				
				Polynomial Feedforward for Linear Parameter-Varying Systems: a Kernel Regularized Approach, \\
				
				To appear in
				
				{\em 22nd IFAC World Congress}, 
				Yokohama, Japan, 2023, 
				
				uploaded to arXiv \today. \\ 
\end{tabular}}}}

\begin{document}
\begin{frontmatter}
\title{
A Kernel-Based Identification Approach to LPV Feedforward: With Application to Motion Systems
}
\thanks[footnoteinfo]{This work is part of the research programme VIDI with project number 15698, which is (partly) financed by the Netherlands Organisation for Scientific Research (NWO). In addition, this research has received funding from the ECSEL Joint Undertaking under grant agreement 101007311 (IMOCO4.E). The Joint Undertaking receives support from the European Union Horizon 2020 research and innovation programme.}
\author[First]{M. van Haren} 
\author[Second,First]{L. Blanken} 
\author[First,Third]{T. Oomen}

\address[First]{Control Systems Technology Section, Eindhoven University of Technology, The Netherlands (e-mail: m.j.v.haren@tue.nl).}
\address[Second]{Sioux Technologies, Eindhoven, the Netherlands}
\address[Third]{Delft Center for Systems and Control, Delft University of Technology, the Netherlands}

\begin{abstract}
The increasing demands for motion control result in a situation where Linear Parameter-Varying (LPV) dynamics have to be taken into account. Inverse-model feedforward control for LPV motion systems is challenging, since the inverse of an LPV system is often dynamically dependent on the scheduling sequence. The aim of this paper is to develop an identification approach that directly identifies dynamically scheduled feedforward controllers for LPV motion systems from data. In this paper, the feedforward controller is parameterized in basis functions, similar to, e.g., mass-acceleration feedforward, and is identified by a kernel-based approach such that the parameter dependency for LPV motion systems is addressed. The resulting feedforward includes dynamic dependence and is learned accurately. The developed framework is validated on an example.
\end{abstract}
\end{frontmatter}

\section{Introduction}
Feedforward control can compensate for known disturbances in motion control, such as a reference trajectory. Typically, a feedforward controller is based on the inverse of a system \citep{Hunt1996,Butterworth2012}, where the control performance is determined by the accuracy of the inverse model \citep{Devasia2002}. The increasing demands for motion control leads to a situation where Linear Parameter-Varying (LPV) dynamics have to be explicitly taken into account \citep{Wassink2005}.

For Linear Time-Invariant (LTI) systems, polynomial feedforward, where the feedforward signal is a linear combination of basis functions, results in good control performance. Often, the basis functions are chosen such that they relate to physical quantities, such as acceleration feedforward for the inertia \Citep{Lambrechts2005,Oomen2019}, and snap feedforward for the compliance of a system \citep{Boerlage2003}. Several approaches have been developed to tune the feedforward parameters based on data, such as iterative learning control \Citep{VanDeWijdeven2010} and instrumental variable identification \citep{Boeren2015} approaches. However, LTI feedforward leads to suboptimal performance when applied to LPV systems.

A key challenge in feedforward for LPV systems is modeling the dependency on the scheduling sequence. Additionally, the inversion of LPV systems generate terms that are often dynamically dependent on the scheduling sequence, i.e., dependency on the derivatives of the scheduling sequence \citep{Sato2003}. Hence, the dependence on the scheduling, including dynamic dependence, should be taken into account for feedforward of LPV systems and directly determines the achievable performance limit.

Several developments have been made in feedforward for LPV systems, and are directed at 1) identification of static LPV feedforward and 2) feedforward techniques based on forward LPV models. Inverse LPV system design is investigated in \citet{Balas2002, Sato2008}, but rely on the forward LPV model and do not take dynamic dependence into account. In \Citet{VanHaren2022a}, position-dependent snap feedforward is developed, that compensates for the static contribution of the position-dependent compliance. Data-driven feedforward approaches are developed in \citet{Butcher2009,DeRozario2018a}, but do not include dynamic dependence. In \citet{Theis2015,DeRozario2017,Bloemers2018}, state-space models of LPV systems are used to create inverse systems and in \citet{Kontaras2016} a compliance compensation is developed, that do include dynamic dependence, but all heavily rely on the quality of the model, which is not addressed in these papers. Hence, current feedforward approaches for LPV systems either do not take dynamic dependence into account, or heavily depend on models, which directly limits the achievable performance and imposes a large burden on modeling effort.

Although feedforward approaches for LPV systems have been substantially developed, techniques for direct and accurate identification of LPV feedforward controllers that include dynamic scheduling dependence, which is required for high-performance motion control, are currently lacking. In this paper, feedforward parameters for a class of LPV motion systems are directly identified using data with kernel-based approaches, see, e.g., \citep{Pillonetto2014,Blanken2020}, which results in a feedforward strategy that includes dynamic dependence on the scheduling sequence, and retains the polynomial feedforward structure, which is often desirable in motion control \citep{Lambrechts2005, Oomen2019}. This relates to the Bayesian approaches in \citep{Golabi2017,Darwish2018}, yet identifies inverse models for feedforward control and enables dynamic dependence on the scheduling sequence. The contributions include
\begin{itemize}
	\item[(C1) ] Development of a feedforward parameterization for LPV motion systems that includes dynamic dependency on the scheduling sequence.
	\item[(C2) ] Identification of feedforward parameters of the developed parameterization by kernel regularized methods.
	\item[(C3) ] Validation of the framework in a benchmark example.
\end{itemize}
The outline in this paper is as follows. In \secRef{sec:problem}, the feedforward problem for LPV motion systems is shown. In \secRef{sec:BFFF}, the developed feedforward parameterization is introduced. In \secRef{sec:kernel}, the identification of LPV feedforward parameters using input-output data is presented. In \secRef{sec:example}, a benchmark example is shown, validating the framework. Finally, in \secRef{sec:conclusions}, a summary and recommendations are given.

\section{Problem Formulation}
\label{sec:problem}
In this section, the problem related to feedforward control for LPV motion systems is formulated. First, the control setting and feedforward goal for LPV motion systems is described. Second, polynomial feedforward for LTI systems is outlined. Third, challenges in designing feedforward controllers for LPV motion systems are shown, that motivate the problem definition in \secRef{sec:pdef}.
\subsection{Control Setting}
\begin{figure}
	\centering
	\includegraphics{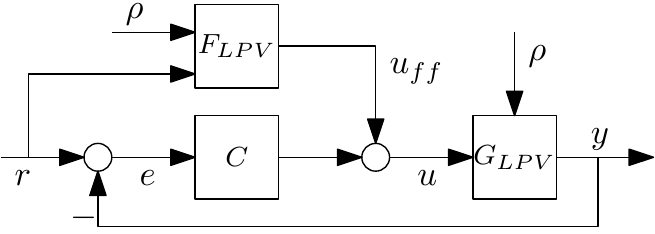}
	\caption{Feedforward structure considered}
	\label{fig:controlStructure}
\end{figure}

The control goal is to develop LPV feedforward controller $F_{LPV}$ to reduce the tracking error $e=r-y$ for single-input single-output LPV system $G_{LPV}$, where perfect tracking is achieved by $F_{LPV}=G_{LPV}^{-1}$. The control structure can be seen in \figRef{fig:controlStructure}, where $C$ is a stabilizing feedback controller.
The reference trajectory $r$ is a smooth reference that can be differentiated at least four times, as in \citet{Lambrechts2005}. The considered class of LPV systems $G_{LPV}$ can be represented in Continuous-Time (CT) by input-output representations and is shown in \defRef{def:LPV}.
\begin{defn}[CT-IO-LPV system]
	\label{def:LPV}
	The considered class of LPV motion systems are statically dependent on the scheduling and given by
	\begin{equation}
		\label{eq:LPVdesc}
			G_{LPV}:\:\sum_{i=-2}^{n_a}a_i(\rho(t)) y^{(i)}(t) \!=\!\sum_{j=0}^{n_{\mathrm{b}}}b_j(\rho(t)) \frac{d^{j}}{d t^{j}} \iint u(t) \, dt^2 
	\end{equation}
with scheduling sequence $\rho \in \mathbb{R}^{n_\rho}$ and $y^{(i)}(t)$ is the ${i}^{\mathrm{th}}$ time derivative of $y(t)$ if $i\geq0$, and the ${i}^{\mathrm{th}}$ integral over time if $i<0$. 
The LPV coefficients $a_i(\rho(t))$ and $b_j(\rho(t))$ have a static dependency on $\rho(t)$, i.e., are not dependent on any derivatives of $\rho$. For ease of notation, the dependence of signals on $(t)$ is from now on omitted.
\end{defn}
The following is assumed of the considered class of LPV systems.
\begin{assum}
	\label{ass:assLPVSystem}
	The following assumption is made for the considered LPV systems.
	\begin{enumerate}
		\item The second integral of the input $u$, i.e., $\iint u \: dt^2$ explicitly appears in the input-output representation of the system.
	\end{enumerate}
This is the case for, e.g., systems where the actuated mass is not connected to the fixed world.
\end{assum}
\begin{rem}
	The LPV coefficients $a_i(\rho)$ and $b_j(\rho)$ have a static dependency on $\rho$, but can be extended to include dynamic dependency, i.e., $a_i(\rho,\dot{\rho},\ldots)$ and $b_j(\rho,\dot{\rho},\ldots)$ with $\dot{\rho}=\frac{d}{dt}\rho$, and is part of ongoing research.
\end{rem}

\subsection{LTI Polynomial Feedforward}
LTI polynomial feedforward, see e.g. \Citet{Lambrechts2005}, approximates an inverse system to reduce the tracking error $e$. 
\begin{defn}[Polynomial feedforward]
	\label{def:polFF}
	Polynomial feedforward is linear in the parameters by approximating the inverse system by neglecting the zero dynamics of the system, i.e., $\sum_{j=1}^{n_b}b_j(\rho(t))\frac{d^j}{dt^j}\iint u(t) \: dt^2$.
\end{defn}
LTI polynomial feedforward applied to the LPV system in \eqref{eq:LPVdesc} parameterizes the feedforward by evaluating \eqref{eq:LPVdesc} at $\rho=\bar{\rho}$, and differentiates both sides twice, resulting in
\begin{equation}
	\label{eq:FFPolLTI}
	F_{LTI}:\quad u_{ff}= \sum_{i=-2}^{n_a}\frac{ a_i(\bar{\rho})}{b_0(\bar{\rho})} \frac{d^{i+2}}{dt^{i+2}}r = \sum_{i=1}^{n_\theta} \theta_i \psi_i\left( \frac{d}{dt}\right)r,
\end{equation}
with differentiators $\psi_i(\frac{d}{dt})$, e.g., $\psi_i(\frac{d}{dt})=\frac{d^2}{dt^2}$ for acceleration feedforward. The parameters can be tuned manually \citep{Lambrechts2005} or estimated with data \citep{Boeren2018a}, that is straightforward due to the linearity in the parameters in \eqref{eq:FFPolLTI}. The resulting feedforward controller is interpretable, simple and effective for LTI systems, but does not take LPV dynamics into account. 
\subsection{Feedforward Problem for LPV Systems}
Developing an inverse model for polynomial feedforward for LPV systems, similar to polynomial feedforward for LTI systems, is challenging due to the dynamic scheduling dependency introduced by deriving an inverse model. The dynamic dependence is observed when inverting \eqref{eq:LPVdesc}, i.e., by differentiating both sides twice, derivatives of the scheduling sequence $\rho$ directly appear. A fixed structure for identifying the inverse model could be used, e.g., a polynomial of $\rho$, $\dot{\rho}=\frac{d}{dt}\rho$ and $\ddot{\rho}=\frac{d^2}{dt^2}\rho$, however, it is unclear how to choose the structure and order. \exampleRef{example:LPVinverse} illustrates the complexity of feedforward control for LPV systems.

\begin{exmp}
	\label{example:LPVinverse}
	(Feedforward problem for LPV system) Consider the two-mass-spring-damper system with parameter-dependent spring in \figRef{fig:LPVMSD}, with input $u$ the force on the first mass, and output $y$ the position of the second mass.
	\begin{figure}
		\centering
		\includegraphics{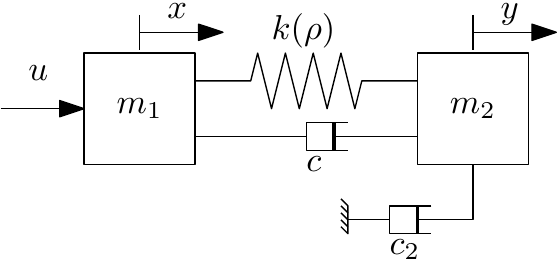}
		\caption{LPV mass-spring-damper with parameter-dependent spring $k(\rho)$ that is considered in this paper.}
		\label{fig:LPVMSD}
	\end{figure}
	The input-output behavior in the form of \eqref{eq:LPVdesc} is given by
	\begin{equation}
		\label{eq:exampleIO}
		\resizebox{\hsize}{!}{%
			$
			\begin{aligned}
				&\left(m_2m_1\frac{d^2}{dt^2} +\left( c\left(m_1+m_2 \right)+c_2m_1\right) \frac{d}{dt}+\left( k(\rho)\left(m_1+m_2 \right)+cc_2 \right)  \right) y \\
				&+k(\rho)c_2 \int y \,dt= \left(c\frac{d}{dt}+k(\rho) \right) \iint u \,dt^2,
			\end{aligned}$}%
	\end{equation}
	which directly shows the double integral of the input signal seen in \eqref{eq:LPVdesc}. LTI polynomial feedforward from \eqref{eq:FFPolLTI} is
	\begin{equation}
		\label{eq:LTIPolFFExample}
		\begin{aligned}
			u_{ff} = &\frac{m_2m_1}{k}\ddddot{r} + \frac{c_2m_1+c\left( m_1+m_2\right)}{k}   \dddot{r}\\
			&+ \left( \left(m_1+m_2 \right)+\frac{cc_2}{k}\right)\ddot{r} + c_2\frac{d}{dt}\dot{r} ,
		\end{aligned}
	\end{equation}
	that consists of the well-known snap, jerk, acceleration and velocity feedforward. However, the true inverse dynamics of \eqref{eq:exampleIO}, when neglecting the zero dynamics of the system, are given by
	\begin{equation}
		\label{eq:exampleIO2}
		\resizebox{\hsize}{!}{
			$
			\begin{aligned}
				u&\approx\frac{m_2m_1}{k(\rho)}\ddddot{y} + \frac{c\left( m_1+m_2\right)+c_2m_1}{k(\rho)} \dddot{y}+ \left(m_1+m_2 \right)\ddot{y} +c_2 \dot{y}\\
				&+ \left( \frac{\frac{2\dot{\rho}^2{k^{\prime}}^2(\rho)}{k(\rho)}-\dot{\rho}^2k^{\prime\prime}(\rho)-\ddot{\rho}k^\prime(\rho)}{k^2(\rho)}-\frac{2 \dot{\rho}k^\prime(\rho)  }{k^2(\rho)}\right)f(y,\dot{y},\ddot{y}),\\
			\end{aligned}$}
	\end{equation}
	with $f(y,\dot{y},\ddot{y}) = m_1m_2  \ddot{y}+\left(c\left( m_1+m_2\right) + c_2m_1\right) \dot{y}\allowbreak+\allowbreak \left(k(\rho)\left( m_1+m_2\right) +cc_2\right)y $, where $k^\prime=\frac{dk(\rho)}{d\rho}$ and $k^{\prime\prime}=\frac{d^2k(\rho)}{d\rho^2}$. When comparing LTI feedforward in \eqref{eq:LTIPolFFExample} with the approximate inverse in \eqref{eq:exampleIO2}, it is observed that LTI feedforward lacks both static and dynamic dependency on $\rho$.
\end{exmp}

\subsection{Problem Definition}
\label{sec:pdef}
A technique for manual tuning or direct data-driven identification of feedforward controllers for LPV motion systems, including dynamic dependency on the scheduling sequence, is currently lacking. The problem addressed in this paper is the direct identification of LPV polynomial feedforward controller of the form
\begin{equation}
	\label{eq:Ftilde}
	\tilde{F}_{LPV}:\quad u_{ff} = \sum_{i=1}^{n_\theta} \theta_i(\rho,\dot{\rho},k^\prime(\rho),\ldots) \psi_i\left( \frac{d}{dt}\right)r,
\end{equation}
based on input-output data $\{u,y\}$, for the class of LPV motion systems in \eqref{eq:LPVdesc}, that includes dynamic dependence on the scheduling sequence $\rho$, where the structure and order of the model for the scheduling dependency is not fixed a priori, and minimizes the tracking error $e$ in the configuration of \figRef{fig:controlStructure}.

\section{Linearly Parameterized Feedforward for LPV motion systems}
\label{sec:BFFF}
In this section, polynomial feedforward strategy for LPV systems as in \eqref{eq:Ftilde} is developed, by posing an alternative parameterization for the LPV motion systems in \eqref{eq:LPVdesc}, that includes dynamic dependency on the scheduling sequence, but simplifies the identification problem significantly.

The key idea is to rewrite the system dynamics in \eqref{eq:LPVdesc} as
\begin{equation}
	\label{eq:IOModel}
	\sum_{i=-n_i}^{n_a}a_i(\rho) y^{(i)}=\sum_{j=0}^{n_{\mathrm{b}}}b_j(\rho) \frac{d^{j}}{d t^{j}} w,
\end{equation}
where a change of variables is used as $w={ \iint} u \, dt^2$.
Similarly to LTI polynomial feedforward in \defRef{def:polFF} and \eqref{eq:FFPolLTI}, feedforward for LPV systems is parameterized by
\begin{subnumcases}{F_{LPV}:}
	w_{ff}=\sum_{i=1}^{n_\theta}\theta_i(\rho) \psi_i\left(\frac{d}{dt},I \right)r, \label{eq:FFModel}
	\\
	u_{ff} = \frac{d^2}{dt^2}w_{ff}, \label{eq:FFCalc}
\end{subnumcases}
where $\psi$ contains differentiators $\frac{d}{dt}$ or integrals $I$, e.g. $\psi_i\left( \frac{d}{dt},I\right) r = \frac{d^2}{dt^2}r=\ddot{r}$ or $\psi_i\left(\frac{d}{dt},I\right) r=Ir = { \int} r \, dt$. 

Note that from \eqref{eq:FFModel}, the second integral of the input $w_{ff}$ is composed out of basis functions, in contrast to the input $u_{ff}$ for LTI polynomial feedforward. The dynamic dependence on the scheduling sequence seen in \eqref{eq:exampleIO2} is introduced by the second derivative with respect to time in \eqref{eq:FFCalc}, which introduces time derivatives of $\theta(\rho)$. In \exampleRef{example:LPVmsdFF}, an example is shown for the two-mass system.
\begin{exmp}[LPV feedforward]
	\label{example:LPVmsdFF}
	Consider the two mass-spring-damper system from \exampleRef{example:LPVinverse}, with input-output behavior in \eqref{eq:exampleIO}. The polynomial feedforward strategy is then defined, by neglecting the zero in the right-hand side of \eqref{eq:exampleIO} according to \assRef{ass:assLPVSystem}, i.e., $\frac{c}{k(\rho)}\approx 0$, as
	\begin{equation}
		\label{eq:exampleFFPara}
		\resizebox{0.85\hsize}{!}{
			$
		\begin{aligned}
			w_{ff} =  &\underbrace{\vphantom{\frac{m_2}{k(\rho)}}c_2}_{\theta_1(\rho)}\underbrace{\vphantom{\frac{m_2}{k(\rho)}}\int}_{\psi_1} r \, dt+\underbrace{\vphantom{\frac{m_2m_1}{k(\rho)}}\left(m_1+m_2 + \frac{cc_2}{k(\rho)} \right)}_{\theta_2(\rho)}\underbrace{\vphantom{\frac{m_2m_1}{k(\rho)}}1}_{\psi_2}r\\
			&+\underbrace{\frac{c\left( m_1+m_2\right)+c_2m_1}{k(\rho)}}_{\theta_3(\rho)}\underbrace{\vphantom{\frac{m_2m_1}{k(\rho)}}\frac{d}{dt}}_{\psi_3}r+\underbrace{\frac{m_2m_1}{k(\rho)}}_{\theta_4(\rho)}\underbrace{\vphantom{\frac{m_2m_1}{k(\rho)}}\frac{d^2}{dt^2}}_{\psi_4}r,
		\end{aligned}$}
	\end{equation}
	where the applied feedforward force is calculated using \eqref{eq:FFCalc}. The applied feedforward force contains both static and dynamic scheduling dependence when substituting \eqref{eq:exampleFFPara} into \eqref{eq:FFCalc}, i.e.,
	\begin{equation}
		\label{eq:exampleResultingFF}
		\resizebox{\hsize}{!}{
			$
		\begin{aligned}
			u_{ff} &= \frac{m_2m_1}{k(\rho)}\ddddot{r} + \frac{c\left( m_1+m_2\right) +c_2m_1}{k(\rho)} \dddot{r}+ \left(m_1+m_2 \right)\ddot{r} +c_2 \dot{r}\\
			& + \left( \frac{\frac{2\dot{\rho}^2{k^{\prime}}^2(\rho)}{k(\rho)}-\dot{\rho}^2k^{\prime\prime}(\rho)-\ddot{\rho}k^\prime(\rho)}{k^2(\rho)}-\frac{2 \dot{\rho}k^\prime(\rho)  }{k^2(\rho)}\right)f(r,\dot{r},\ddot{r}),\\
		\end{aligned}
	$}
	\end{equation}
which is equal to \eqref{eq:exampleIO2} when substituting $y$ for $r$.
\end{exmp}
The applied feedforward force $u_{ff}$ in \eqref{eq:FFCalc} includes the dynamic dependency on the scheduling signal, e.g. shown in \eqref{eq:exampleIO2}, while the modeled $w_{ff}$ in \eqref{eq:FFModel} is only statically dependent on the scheduling sequence.


\section{Kernel Regularized Learning of LPV Feedforward Parameters}
\label{sec:kernel}
In this section, the functions $\theta_i(\rho)$ in \eqref{eq:FFModel} are identified using kernel regularization, which models the functions without a specified structure or order, since the solution is in the infinite-dimensional Reproducing Kernel Hilbert Space (RKHS). Second, kernel design for LPV feedforward parameters is described. Finally, the developed approach is summarized in a procedure.
\subsection{Kernel Regularized Identification}
Given a system $G_{LPV}:\:u\mapsto y$, a model mapping $y$ to $w=\iint u \: dt^2$ is to be identified. A cost function is defined using input-output data as \citep{Pillonetto2014,Blanken2020}
\begin{equation}
	\label{eq:minimization}
	\hat{\Theta} = \arg \min_\Theta \left\| \overline{w}-\Phi\Theta\right\|^2 + \gamma\|\Theta\|^2_\mathcal{H},
\end{equation}
with Euclidean norm $\|\cdot\|$, $\Phi\Theta$ equal to $w_{ff}$ in \eqref{eq:FFModel}, and measurement data vector $\overline{w}$, that is constructed as
\begin{equation}
	\label{eq:defineWbar}
	\overline{w}= \begin{bmatrix}w(0T_s) & w(1T_s) & \cdots & w((N-1)T_s)\end{bmatrix}^\top.
\end{equation}
The squared induced norm on the RKHS $\mathcal{H}$ is denoted as $\|\Theta\|^2_\mathcal{H}$, that is given by \citep{Pillonetto2014},
\begin{equation}
	\label{eq:RKHS}
	\|\Theta\|^2_\mathcal{H} = \Theta^\top K^{-1} \Theta,
\end{equation}
with kernel $K$. The parameter vector $\Theta$ and basis function matrix $\Phi$ are built up as
\begin{equation}
	\label{eq:ThetaPhi}
	\begin{aligned}
		\Theta = \begin{bmatrix}
			\overline{\theta}_1^\top &
			\overline{\theta}_2^\top &
			\cdots &
			\overline{\theta}_{n_\theta}^\top
		\end{bmatrix}^\top, &&
		\Phi = \begin{bmatrix}
			\overline{\phi}_1 & \overline{\phi}_2 & \cdots & \overline{\phi}_{n_\theta}
		\end{bmatrix},
	\end{aligned}
\end{equation}
where the individual parameter vector $\overline{\theta}_i$ and $\overline{\phi}$ are constructed by gathering the values over a training period as
\begin{equation}
	\label{eq:defineThetaPhibar}
			\resizebox{0.85\hsize}{!}{
		$
	\begin{aligned}
		\overline{\theta}_i &= \begin{bmatrix}
			\left( \theta_i(\rho)\right)\left( 0T_s\right)  \\
			\left(\theta_i(\rho)\right)\left( 1T_s\right)  \\
			\vdots \\
			\left(\theta_i(\rho)\right)\left( \left(N-1 \right) T_s)\right) 	\end{bmatrix} \\
		\overline{\phi}_{i}&=\begin{bmatrix}
			\left(\psi_{i} y\right)\left(0 T_{s}\right) & 0 & \cdots & 0 \\
			0 & \left(\psi_{i} y\right)\left(1 T_{s}\right) & \cdots & \vdots \\
			\vdots & \ddots & \ddots & \vdots \\
			0 & \cdots & \cdots & \left(\psi_{i} y\right)\left((N-1) T_{s}\right)
		\end{bmatrix}
	\end{aligned}
$}
\end{equation}
where $(\frac{d}{dt},I)$ has been left out for brevity. 
\begin{rem}
	Note that the calculation of $\bar{\phi}_i$ \eqref{eq:defineThetaPhibar} might require taking the derivative of the output. In the presence of noise, this can be done by, e.g., using Kalman estimation or low-pass filtering. A framework where no derivatives of the output are used is part of ongoing research.
\end{rem}
The solution to the cost function in \eqref{eq:minimization} is given by \citep{Pillonetto2014}
\begin{equation}
	\label{eq:solTheta}
	\hat{\Theta} =K \Phi^{\top}\left(\Phi K \Phi^{\top}+\gamma I_{N}\right)^{-1} \overline{w},
\end{equation}
where parameters $\theta$ are estimated at any $\rho^*$ using the representer theorem \cite[Section~9.2]{Pillonetto2014}.
 \begin{rem}
	\label{rem:bias}
Note that \eqref{eq:minimization} is an open-loop solution, while a closed-loop control structure is assumed as shown in \figRef{fig:controlStructure}, hence measurement noise introduces bias \Citep{Blanken2020}. The addition of instrumental variables is capable of removing this bias, and is reported elsewhere.
\end{rem}
The kernel $K$ can be designed to incorporate prior knowledge on the feedforward parameters, such as smoothness or periodicity, and will be discussed in the next section.

\subsection{Kernels for LPV Feedforward Parameters}
The kernel incorporates prior knowledge on the feedforward parameters, hence is important to design carefully. The optimal kernel for solving \eqref{eq:minimization} and minimizing the mean-squared error \Citep{Pillonetto2014}, when treating feedforward parameters as random variables, is equal to
\begin{equation}
	\label{eq:optKernel}
		\resizebox{0.88\hsize}{!}{
		$
	\Pi = \mathbb{E}\left(\Theta \Theta^\top \right) = \begin{bmatrix}
		\mathbb{E}( \overline{\theta}_1\overline{\theta}_1^\top) & 	\mathbb{E}( \overline{\theta}_1\overline{\theta}_2^\top) & 	\cdots & 	\mathbb{E}( \overline{\theta}_1\overline{\theta}_{n_\theta}^\top) \\
		\mathbb{E}( \overline{\theta}_2\overline{\theta}_1^\top) & \mathbb{E}( \overline{\theta}_2\overline{\theta}_2^\top) & \cdots & \vdots \\
		\vdots & \vdots & \ddots & \vdots \\
		\mathbb{E}( \overline{\theta}_{n_\theta}\overline{\theta}_1^\top) & \cdots & \cdots & \mathbb{E}( \overline{\theta}_{n_\theta}\overline{\theta}_{n_\theta}^\top)
	\end{bmatrix}. $}
\end{equation}
For LPV motion systems, parameters may correlate, i.e., $\mathbb{E}\left( \overline{\theta}_i\overline{\theta}_j\right) \neq 0 \, \forall j\neq i$. For example, when looking at \eqref{eq:exampleFFPara}, parameters $\theta_3$ and $\theta_4$ are scaled versions of each other. Hence, the framework is capable of incorporating correlation between feedforward parameters.

The optimal kernel is approximated by a kernel matrix,
\begin{equation}
	\label{eq:covker}
	\mathbb{E}\left( \overline{\theta}_i\overline{\theta}_j^\top\right) = K_{ij}(\overline{\rho},\overline{\rho}),
\end{equation}
which only has a static dependency on $\rho$, while the framework produces feedforward which is dynamically dependent on the scheduling sequence as shown in \secRef{sec:BFFF}. 

The kernel matrix $K_{ij}$ is determined by evaluating a kernel function, such as the squared exponential kernel function 
\begin{equation}
	\label{eq:SEKernel}
	k_{ij,SE}(\rho,\rho^\prime) = \sigma_{ij}^2\exp\left( -\frac{\left( \rho-\rho^\prime\right) ^2}{2\ell_{ij}^2}\right) .
\end{equation}
The hyperparameters of the kernel, i.e., for the squared exponential kernel in \eqref{eq:SEKernel} the output variances $\sigma_{ij}^2$ and length scales $\ell_{ij}$ can be tuned using marginal-likelihood optimization. The kernel choice provides the user to apply prior knowledge on the feedforward parameters.

\subsection{Developed Procedure}
The developed procedure is summarized in Procedure~\ref{proc:1}.
\vspace{3pt}\hrule\begin{proced} \textit{(Kernel regularized LPV feedforward identification)} \hfill \vspace{0.5mm} \hrule
	\label{proc:1}
	\begin{enumerate}
		\item Apply reference $r$ to closed-loop system in \figRef{fig:controlStructure} and record $y$, $\rho$ and $u$.
		\item Construct kernel matrix $K$, e.g., based on prior expectations on parameters $\theta_i$.
		\item Calculate matrix $\Phi$ using \eqref{eq:ThetaPhi} and \eqref{eq:defineThetaPhibar}.
		\item Compute $\bar{w}$ from \eqref{eq:defineWbar} using the second integral of the input $w=\iint u \: dt^2$.
		\item Estimate the feedforward parameters $\hat{\Theta}$ using \eqref{eq:solTheta}.
	\end{enumerate}
	\vspace{0pt} 	\hrule 	\vspace{-2pt}
\end{proced}
To conclude, kernel regularized identification is capable of identifying LPV feedforward parameters with input-output data of a system, without specifying a structure. In the following section, an example is shown that validates the developed framework.
\section{Example}
\label{sec:example}
In this section, the developed approach of feedforward for LPV systems is validated on an example.
\subsection{Example Setup}
The two-mass-spring-damper in \figRef{fig:LPVMSD} is considered. The feedback controller $C$ is a lead filter. The system is seen in \eqref{eq:exampleIO} and \figRef{fig:LPVMSD}, with damper constants $c=1$ and $c_2=10^{-4}$ Ns/m, masses $m_1=1$ and $m_2=0.5$ kg. Stiffness $k(\rho)$ is 
\begin{equation}
	k(\rho) = \frac{EA}{\rho(L-\rho)},
\end{equation}
with length $L=1$ m, Young's modulus $E=0.24\cdot10^9$ Pa and area $A=1\cdot10^{-5}\text{ m}^2$. The reference $r$ is chosen as a fourth order point-to-point motion, as in \citet{Lambrechts2005}, consisting of 1810 samples. The scheduling sequence $\rho$ used is the reference itself, and ranges from 0.2 m to 0.8 m. The feedforward parameters $\theta_i(\rho)$ in \eqref{eq:FFModel} are identified according to Procedure~\ref{proc:1}.
\subsection{Compared Approaches}
The following three approaches are compared in feedforward to evaluate the developed framework.
\begin{description}
	\item[LTI] Completely ignoring the LPV dynamics of the system and using static feedforward parameters as in \eqref{eq:FFPolLTI}, where the feedforward parameters are taken as the true parameters for $\rho=0.5$ m, i.e.
	\begin{equation}
		\label{eq:LTIFFsim}
		F_{LTI}: \quad u_{ff} = \theta_1 \dot{r} + \theta_2 \ddot{r} + \theta_3 \ddddot{r}.
	\end{equation}
	\item[Static LPV] Including the LPV effects in the standard polynomial snap feedforward, but ignoring the additional terms which arise due to the chain and product rule of integration \Citep{VanHaren2022a}, i.e.,
	\begin{equation}
		\label{eq:staticLPVFF}
		F_{SLPV}: \quad u_{ff}  = \theta_1 \dot{r} + \theta_2 \ddot{r} + \theta_3(\rho) \ddddot{r}.
	\end{equation}
	\item[Dynamic LPV] Application of the developed \mbox{feedforward} approach in \eqref{eq:FFModel} and \eqref{eq:FFCalc} with dynamic snap feedforward, i.e.,
	\begin{subnumcases}{F_{LPV}:}
		w_{ff}= \theta_1 \int {r} \, dt + \theta_2 {r} + \theta_3(\rho) \ddot{r}., \label{eq:FFModelSim}
		\\
		u_{ff} = \theta_1 \dot{r} + \theta_2 \ddot{r} + \theta_3(\rho) \ddddot{r}+u_{dyn}, \label{eq:FFCalcSim}
	\end{subnumcases}
with 
\begin{equation}
	\label{eq:udyn}
	u_{dyn} = \ddot{\rho}\theta^\prime_3(\rho)\ddot{r}+\dot{\rho}^2\theta_3^{\prime\prime}(\rho)\ddot{r}+2\dot{\rho}\theta_3^\prime(\rho)\dddot{r}.
\end{equation}
\end{description}
The parameters $\theta_i(\rho)$ are identified using the developed framework, where, for simplicity,  the kernel is chosen block-diagonal, i.e., $K_{ij}=0 \, \forall i\neq j$, meaning different feedforward parameters $\theta_i$ and $\theta_j \, \forall i\neq j$ are not expected to correlate. The kernels $K_{11}$ and $K_{22}$ are chosen to be identity matrices of appropriate size, i.e., parameters $\theta_1$ and $\theta_2$ are constant. The kernel $K_{33}$ is chosen as the squared exponential kernel \eqref{eq:SEKernel}, where $\sigma_{33}^2$ and $\ell_{33}$ are optimized using marginal likelihood optimization.

\subsection{Results} 
\begin{figure}[tb]
	\centering
	\hspace{-15mm}
	\includegraphics{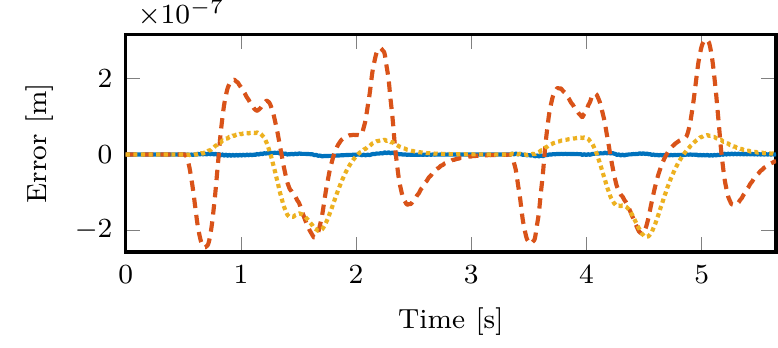}
	\caption{Time-domain error for LTI \li{red}{densely dashed}[1.5], static LPV \li{yel}{densely dotted}[1.5] and developed \li{blue}{solid}[1.5] feedforward approaches.}
	\label{fig:timeDomainError}
\end{figure}

\begin{figure}[tb]
	\centering
	\includegraphics{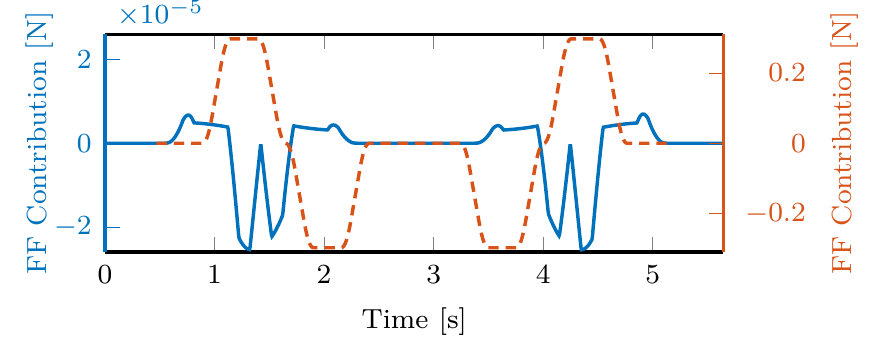}
	\caption{Left axis: Dynamic dependent feedforward $u_{dyn}$ from \eqref{eq:udyn} \li{blue}{solid}[1.5]\hspace{-1mm}. Right axis: Total feedforward \li{red}{densely dashed}[1.5]\hspace{-1mm}.}
	\label{fig:dynContribution}
\end{figure}

\begin{figure}[tb]
	\centering
	\hspace{-15mm}
	\includegraphics{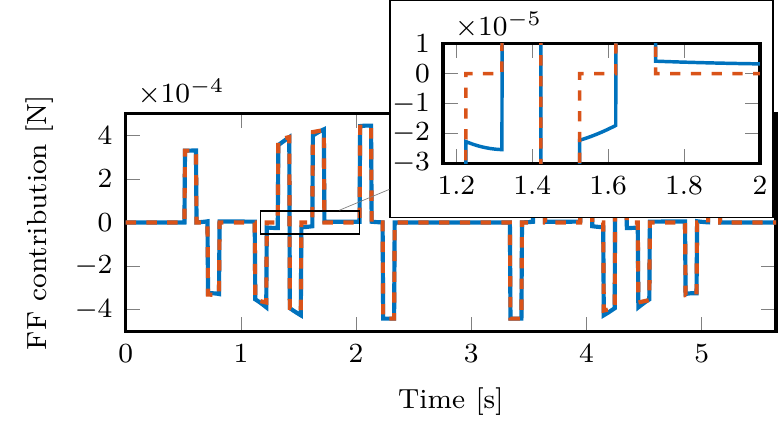}
	\caption{Feedforward contribution of static LPV $\theta_3(\rho)\ddddot{r}$ \li{red}{densely dashed}[1.5] and developed approach $\frac{d^2}{dt^2}\left(\theta_3(\rho) \ddot{r}\right) $\li{blue}{solid}[1.5]. Note that the difference is solely caused by the dynamic scheduling dependency.}
	\label{fig:staticvsdyamic}
\end{figure}

\begin{figure}[tb]
	\centering
	\includegraphics{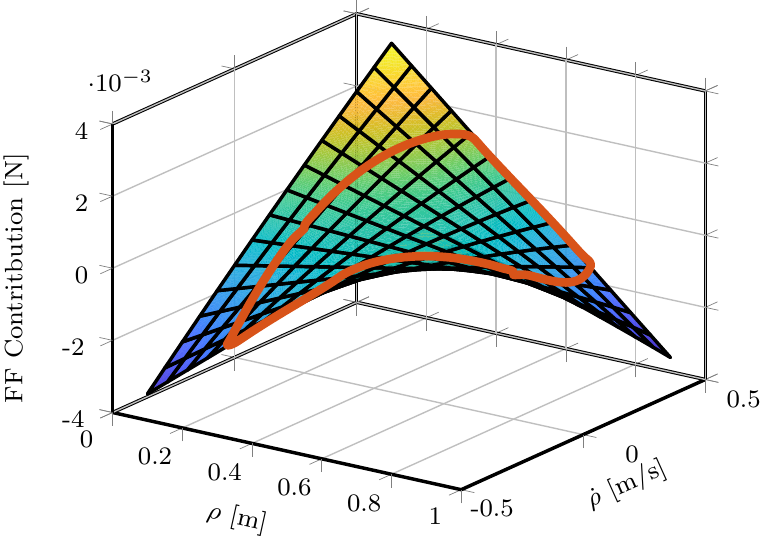}
	\vspace{-1mm}
	\caption{True dynamic contribution $u_{dyn}$ from \eqref{eq:udyn} and estimated given the data \li{red}{solid}[2] for $\ddot{r}=1$ and $\dddot{r}=20$.}
	\vspace{-1mm}
	\label{fig:surface}
\end{figure}

In this section, the results of the example are shown. In \figRef{fig:timeDomainError}, an error plot is shown for the three feedforward approaches. 
In \figRef{fig:dynContribution}, the contribution of the developed feedforward approach due to the dynamic dependency is shown. The contribution of snap feedforward for both static LPV and the developed feedforward approach is shown in \figRef{fig:staticvsdyamic}. A surface plot of the true and estimated dynamic dependent snap feedforward is shown in \figRef{fig:surface}. The following observations are made:
\begin{itemize}
	\item \figRef{fig:timeDomainError} shows that the best tracking performance is achieved by the developed approach, while the static LPV feedforward performs better than LTI feedforward. The root-mean-square errors are respectively $1.4\cdot10^{-9}$ m,  $5.9\cdot10^{-8}$ m and $9.9\cdot10^{-8}$ m.
	\item \figRef{fig:dynContribution} and \figRef{fig:staticvsdyamic} show the dynamic contribution $u_{dyn}$ to the feedforward, which explains the performance difference between static LPV feedforward and the developed dynamic LPV feedforward.
	\item For a high contribution of the dynamic feedforward in \figRef{fig:dynContribution}, the tracking error for static LPV in \figRef{fig:timeDomainError} increases, showing that the dynamic dependence has significant effect on the tracking error.
	\item \figRef{fig:surface} shows that, for the reference, the dynamic contribution to the feedforward is estimated accurately.
\end{itemize}

\section{Conclusions}
\label{sec:conclusions}
In this paper, a method is developed to directly identify feedforward controllers for LPV motions systems, including dynamic dependence on the scheduling sequence. A polynomial feedforward model is developed for LPV motion systems by using a change of variables, i.e., the double integral of the input signal. The feedforward parameters are directly identified with input-output data using a kernel-regularized approach. An example shows that tracking performance is significantly improved compared to existing LTI or LPV approaches that do not take dynamic dependence into account. \\
Ongoing research is aimed at adding instrumental variables and directly learning LPV feedforward parameters without change of variables. Finally, extension to a broader feedforward structure and experimental validation of the framework is part of ongoing work.

\bibliographystyle{ifacconf}
\bibliography{library}


\end{document}